\documentclass[%
reprint,
superscriptaddress,
amsmath,amssymb,
aps,
l
]{revtex4-1}

\usepackage{color}
\usepackage{graphicx}
\usepackage{dcolumn}
\usepackage{bm}

\begin {document}

\title{
Distributional Behaviors of Time-averaged Observables 
in Langevin Equation with Fluctuating  Diffusivity: Normal Diffusion but Anomalous Fluctuations}

\author{Takuma Akimoto}
\email{akimoto@keio.jp}
\affiliation{%
  Department of Mechanical Engineering, Keio University, Yokohama, 223-8522, Japan
}%

\author{Eiji Yamamoto}
\affiliation{%
  Department of Mechanical Engineering, Keio University, Yokohama, 223-8522, Japan
}%

\date{\today}

\begin{abstract}
We consider Langevin equation with dichotomously fluctuating diffusivity, where the diffusion coefficient changes dichotomously in time, in order to study fluctuations of time-averaged observables in temporary heterogeneous diffusion 
process.  
We find that occupation time statistics is a powerful tool for calculating the time-averaged mean square displacement 
in the model. We show that  the time-averaged diffusion coefficients are intrinsically random when the mean sojourn time for 
one of the states diverges. 
Our model provides anomalous fluctuations of time-averaged diffusivity, which have relevance to large fluctuations 
of the diffusion coefficient in single-particle-tracking experiments.
\end{abstract}

\maketitle


\section {Introduction}
Law of large numbers plays an important role in statistical physics. 
In stationary stochastic processes $\bm{X}_{t}$, law of large numbers or the
central limit theorem tells us that time-averaged observables such as
diffusivity and the ratio of occupation time converge to a constant when the
measurement time goes to infinity:
\begin{equation}
\int_0^t O( \bm{X}_{t'})dt'/t \to \langle O(\bm{X}) \rangle ~{\rm as}~ t\to \infty, 
\end{equation}
where the observable $O(\cdot)$ is a function of the stochastic process
$\bm{X}_{t}$. 
In experiments, time-averaged observables are not constant because of
finite measurement times. However, in some stochastic processes describing 
non-equilibrium phenomena, 
time-averaged observables are intrinsically random because of the breakdown of
law of large numbers or the central limit theorem \cite{Darling1957, Lamperti1958}.  In other words, they do not converge to a constant even when
the measurement time goes to infinity and the fluctuations never disappear.  
Such anomalous behavior has been studied
by infinite ergodic theory in dynamical systems \cite{Aaronson1997}.  Infinite
ergodic theory states that time-averaged observables converge in distribution,
and the distribution function depends on the invariant measure as well as a
class of the observation function \cite{Aaronson1981, Thaler1998, Akimoto2008,
  Akimoto2015}.

Continuous-time random walk (CTRW) is a model of anomalous diffusion, where the
mean square displacement (MSD) increases sub-linearly with time, and is extensively 
studied in disorder materials \cite{Scher1975} as well as biophysics \cite{He2008, Weigel2011}. 
In CTRW, a
random walker waits for the next jump and the waiting time is a random
variable whose probability density function (PDF) $\rho(\tau)$ follows a
power-law distribution:
\begin{equation}
\rho(\tau) \sim \frac{c_0}{|\Gamma(-\alpha)|}  \tau^{-1-\alpha}\quad (\tau \to \infty),
\label{power-law-pdf}
\end{equation}
where $c_0$ is a scale factor. When $\alpha \leq 1$, the mean waiting time diverges, thereby
causing a breakdown of law of large numbers and the central limit
theorem. In this case, it was shown that the time-averaged MSD (TMSD) for a fixed lag time $\Delta \ll t$, defined as
\begin{equation}
 \label{tamsd_definition}
  \overline{\delta^{2}(\Delta;t)} \equiv
  \frac{1}{t - \Delta} \int_{0}^{t - \Delta} dt' \,
  [\bm{r}(t' + \Delta) - \bm{r}(t')]^{2} ,
\end{equation}
does not converge to a constant but converges in distribution 
 as $t\to\infty$ \cite{Lubelski2008, He2008, Akimoto2010}.  Moreover, the
PDF of the normalized TMSD, i.e., $\overline{\delta^{2}(\Delta;t)}/ \langle
\overline{\delta^{2}(\Delta;t)} \rangle$, follows a universal distribution
called the Mittag-Leffler distribution, which is one of distributional limit
theorems in infinite ergodic theory \cite{Akimoto2010}.  This distributional
property for a time-averaged observable is called distributional ergodicity in
stochastic processes \cite{Miyaguchi2011a, Akimoto2013a}.

Other distributional behaviors have been found in other diffusion processes such
as a quenched trap model \cite{Miyaguchi2011, Miyaguchi2015} and
stored-energy-driven Levy flight (SEDLF) \cite{Akimoto2013a, Akimoto2014}, where the PDF
of the normalized TMSDs (time-averaged diffusion coefficients) follows other
distributions depending on the power-law exponent in the waiting time distribution,
the spatial dimension as well as parameters controlling jumps of a random
walker.  It is important to clarify whether fluctuations of
time-averaged observables are intrinsic or not, because diffusion coefficients
obtained by single-particle-tracking experiments in living cells exhibit large
fluctuations \cite{Golding2006, Jeon2011, Weigel2011, Tabei2013, Manzo2015}. 
Such large
fluctuations will have relevance to distributional behaviors in stochastic
models of anomalous diffusion.

\section {Langevin equation with dichotomously fluctuating Diffusivity}

To investigate ergodic properties in heterogeneous diffusion processes, we
consider the following Langevin equation with fluctuating diffusivity (LEFD),
\begin{equation}
  \label{e.lefd}
 \frac{d\bm{r}(t)}{dt} = \sqrt{2 D(t)} \bm{w}(t),
\end{equation}
where $\bm{w}(t)$ is the $n$-dimensional white Gaussian noise with
  $\left\langle \bm{w}(t) \right\rangle = 0$, and 
  $\left\langle w_i(t)w_j(t') \right\rangle = \delta_{ij}\delta(t-t')$.
On the other hand, the diffusion coefficient $D(t)$ can be a non-Markovian
stochastic process. We assume that $D(t)$ and $\bm{w}(t)$ are
statistically independent. Because the diffusion coefficient is determined by 
shape of the particle or surrounding environment, the LEFD can describe the dynamics of a particle with inner degree 
of freedom. In fact, 
this model can be utilized in the equation of motion
for the center-of-mass of entangled polymer in reptation model
\cite{Doi-Edwards-book} and is related to dynamic heterogeneity in supercooled
liquids \cite{Yamamoto-Onuki-1998, Yamamoto-Onuki-1998a, Richert-2002, Helfferich2014}.
Moreover, because the stochastic process $D(t)$ is generic, this system includes 
temporally heterogeneous diffusion models induced by spacial heterogeneity 
such as the ones studied in \cite{Massignan-Manzo-TorrenoPina-GarciaParajo-Lewenstein-Lapeyre-2014,Chubynsky2014,Akimoto2015b}

\begin{figure}
\includegraphics[width=.8\linewidth, angle=0]{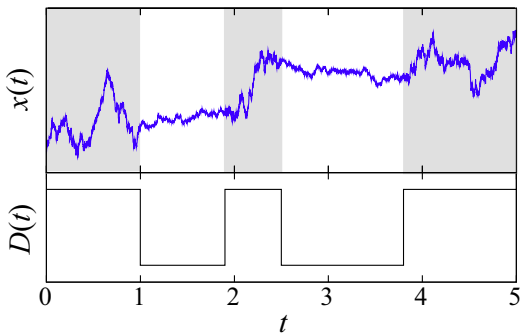}
\caption{Trajectory of the Langevin equation with dichotomously fluctuating diffusivity. 
The lower inset represents the underlying diffusion coefficient. 
 }
\label{traj_D}
\end{figure}


In our previous study \cite{Uneyama2015}, we have obtained the relative standard deviation (RSD) of
the TMSD as a function of measurement time $t$
in LEFD when the stochastic process $D(t)$ is in equilibrium, where the RSD is
defined by
\begin{equation}
  \label{rsd_square_def}
  \Sigma(t;\Delta) \equiv \frac{\sqrt{\langle [\overline{\delta^{2}(\Delta;t)} -
  \langle \overline{\delta^{2}(\Delta;t)} \rangle]^{2} \rangle}}{\langle \overline{\delta^{2}(\Delta;t)} \rangle} .
\end{equation}
In equilibrium processes, the RSD becomes 
\begin{equation}
 \label{rsd_square_final}
  \Sigma^{2}(t;\Delta)
  \approx \frac{2}{t^{2}} \int_{0}^{t} ds 
  (t - s) \psi_1(s),
\end{equation}
where $\psi_1(t)$ is the normalized correlation function of diffusion
coefficients, i.e., $\psi_1(t) \equiv (\langle D(t) D(0)
\rangle - \langle D \rangle^2)/\langle D
\rangle^2$. Therefore, information on the underlying diffusion coefficient 
$D(t)$ can be extracted by the RSD analysis \cite{Uneyama2012,Uneyama2015}.
Here, we investigate ergodic properties of LEFD especially in non-equilibrium cases. In
particular, we consider two-state models for the stochastic process $D(t)$. When
the mean sojourn time of a state in $D(t)$ diverges, the stochastic process
becomes non-stationary, which implies that the system is intrinsically in
non-equilibrium. 
We show normal diffusion yet anomalous fluctuations of TMSD. 

Here, we consider dichotomous processes for diffusivity $D(t)$ (see Fig.~1), i.e., 
$D(t)=D_+$ if the state is $+$ and $D(t)=D_-$ otherwise ($-$ state). 
Sojourn times for $+$ and $-$ states are random variables following different
probability density functions (PDFs), $\rho_+(\tau)$ and $\rho_-(\tau)$ for $+$ and
$-$ states, respectively. We assume that the one of the PDFs $\rho_+(\tau)$ follows either a narrow distribution
where all moments are finite or a broad distribution of power-law form [Eq.~(\ref{power-law-pdf})],
and that the other PDF follows a power-law distribution, whose Laplace
transform is given by 
  $\hat{\rho}_-(s) = 1 - a_- s^{\alpha_-}   + o(s)$ 
  ($\alpha_-<1$).
  In particular, we consider three cases for $\rho_+(x)$:
  	\begin{itemize}
    	\setlength{\leftskip}{5.5mm}
    	\item [(1)] narrow distribution: \hspace*{.5cm}
    	$\hat{\rho}_+(s) = \sum_{k=0}^\infty \frac{m_k}{k!} s^k$,\\[-.2cm]
    	\item [(2)] $\alpha_-<\alpha_+ <1$: \hspace*{0.25cm}
    	$\hat{\rho}_+(s) = 1 - a_+ s^{\alpha_+} + o(s^{\alpha_+})$,\\[-.2cm]
    	\item [(3)] $\alpha_- = \alpha_+$: \hspace*{0.9cm}
    	$\hat{\rho}_+(s) = 1 - a_+ s^{\alpha_+} + o(s^{\alpha_+})$,
  	\end{itemize}
where $m_k$ is the $k$th moment of  sojourn times of the state $+$. 
In what follows, we set $\alpha_-=\alpha$. This kind of power-law behavior is observed in supercooled 
liquids \cite{Helfferich2014}. 

\section{ Representation of time-averaged mean square displacement}
For $\Delta \ll t$, TMSD is represented by 
\begin{align}
  \label{e.0}
  \overline{\delta^{2}(\Delta;t)} 
  \underset{\Delta \ll t}{\approx}\!
  \frac{\displaystyle
    \sum_{i=0}^{N_t-1}\!\!
    \int_{t_i}^{t_{i+1}} \!\!\!\!\!\!\delta \bm{r}^2(\Delta;t') dt' +\!
    \int_{t_{N_t}}^{t}\!\!\!\!\! \delta \bm{r}^2(\Delta;t') dt'}{t },
\end{align}
where $\delta \bm{r}(\Delta;t')\equiv \bm{r}(t'+\Delta) - \bm{r}(t')$, $t_i$ is
the $i$th transition time from one state to the other state with $t_0=0$, $N_t$
is the number of transitions up to time $t$. Since a particle undergoes Brownian motion in each state, 
\begin{align}
  \sum_{i=0}^{N_t-1}&\int_{t_i}^{t_{i+1}} \delta \bm{r}^2(\Delta;t') dt' + \int_{t_{N_t}}^{t} \delta \bm{r}^2(\Delta;t') dt'\notag\\
  \label{e.3}
  \underset{\Delta\ll \tau_0}{\approx}&
  \int_{0}^{T_+(t_{N_t})} \!\!\!\delta \bm{r}_+^2 (\Delta; t') dt'
  +\!
  \int_{0}^{T_-(t_{N_t})} \!\!\!\delta \bm{r}_-^2 (\Delta; t') dt',
\end{align}
where $\delta \bm{r}_{\pm}(\Delta; t') \equiv \int_{t'}^{t'+\Delta} dt''
\sqrt{2D_{\pm}} \bm{w}(t'')$, $T_{\pm}(t)$ is the occupation time of the
state $\pm$ up to time $t$ [Thus, $T_+(t) + T_-(t) = t$], and $\tau_0$ is a characteristic time 
for the transitions of $D(t)$. The condition of $\Delta \ll \tau_0$ validates the approximation 
 that the state in $[t_i, t_i + \Delta]$ does not change. 
%
We have
\begin{align}
  \overline{\delta^{2}(\Delta;t)}
  &\approx
  \label{e.tmsd.approx}
  2n 
  \frac{
    \overline{D_+(t)}\, T_+(t) +
    \overline{D_-(t)}\, T_-(t)}{t} \Delta,
\end{align}
where we define a time-averaged diffusion coefficient of each state as
$
\overline{D_\pm(t)} \equiv 
\int_{0}^{T_\pm(t) } \delta \bm{r}^2_\pm(\Delta;t') dt'/2nT_\pm(t).  
$
Therefore, TMSDs always show normal diffusion and the time-averaged diffusion
coefficient defined as $\overline{D(t)}\equiv \overline{\delta^{2}(\Delta;t)}/(2n\Delta)$ is given by
\begin{equation}
  \overline{D(t)} \approx
  \overline{D_-(t)}  +
  \left[\overline{D_+(t)} - \overline{D_-(t)}\right]\frac{T_+(t)}{t}.
  \label{D_t_TAMSD}
\end{equation}
Using Eqs.~(\ref{e.tmsd.approx}) and (\ref{D_t_TAMSD}), we have the RSD
[Eq.~(\ref{rsd_square_def})]:
$
  \Sigma^2(t; \Delta)
  \approx
  \langle [\overline{D(t)} -
      \langle \overline{D(t)} \rangle ]^{2} \rangle /
  \langle \overline{D(t)} \rangle^2 .
$

In Eq.~(\ref{D_t_TAMSD}), the time-averaged diffusion coefficient
$\overline{D(t)}$ is controlled by three stochastic variables,
$\overline{D_{\pm}(t)}$, and $T_+(t)$. As shown below, the RSD of $T_+(t)$
decays slowly $t^{-\beta}$ with $\beta < 1/2$ in the limit $t \to \infty$, while 
those of $\overline{D_{\pm}(t)}$ decay as $t^{-0.5}$. Therefore, in the long
time limit, the fluctuations of $T_+(t)$ is dominant over those of
$\overline{D_{\pm}(t)}$, and thus we can approximate as $\overline{D_\pm(t)}
\simeq D_\pm$. Under this approximation, we have an asymptotic behavior of the RSD:
\begin{equation}
\Sigma^{2} (t; \Delta) \sim 
\frac{
    \left\langle T_+^2(t) \right\rangle  -
    \left\langle T_+(t)   \right\rangle^2
  }
  {\frac {D_-}{(D_+ - D_-)^2}t^2 +  \langle T_+(t) \rangle^2}.
\label{rsd_occupation_time}
\end{equation}
This is another representation of the RSD by the occupation time in LEFD with two-state diffusivity. We confirmed that 
 the asymptotic behavior is the same as the RSD (\ref{rsd_square_final}) in equilibrium processes. 
Since we neglect fluctuations of $\overline{D_\pm (t)}$, this expression for the RSD is valid only when the right-hand side of 
Eq.~(\ref{rsd_occupation_time}) decays slower than $t^{-0.5}$. Otherwise, the asymptotic behavior of the RSD is the same 
as that in Brownian motion (see Appendix.~A): 
\begin{equation}
\Sigma^{2} (t; \Delta) \sim \frac{\langle (\overline{D_-(t)}-D_-)^2\rangle}{D_-^2} \sim \frac{4\Delta }{3nt}.
\label{rsd_occupation_time2}
\end{equation}

\section{ Occupation time statics} 
Here, we consider the occupation time statics for three cases. 
We define the joint probability
distribution, $g^{\pm}_n(y;t)$, of the occupation time $T_+(t)=y$ and the number
of renewal $N_t = n$ up to time $t$ under the condition that the initial state is
$\pm$, given by
\begin{equation}
  g^{\pm}_n(y;t) =
  \left\langle
  \delta\left(y-T_+(t)\right) I\left(t_n\leq t<t_{n+1}\right)
  \right\rangle_\pm.
\end{equation}
The Laplace transform of $g^{\pm}_n(y;t)$ with respect to $y$ and $t$ is given
by
\begin{equation}
  \hat{g}^{\pm}_n(u;s)
  = \left\langle \int_{t_n}^{t_{n+1}} e^{-st} e^{-u T_+(t)} dt\right\rangle_{\pm},
\end{equation}
where $n=1,2,\dots$. For example, if the initial state is $+$ and $n=2k$ or
$2k+1$, it can be represented as
\begin{align}
  \hat{g}^{\pm}_{2k}(u;s)
  &=
  \left\langle
  \int_{t_{2k}}^{t_{2k+1}} \!\!\! e^{-st} e^{-u [\tau_1 + \tau_3 + \dots + \tau_{2k-1}+ (t-t_{2k})]}dt
  \right\rangle,
  \\[-0.1cm]
  \hat{g}^{\pm}_{2k+1}(u;s)
  &=
  \left\langle
  \int_{t_{2k+1}}^{t_{2k+2}}\!\!\! e^{-st} e^{-u (\tau_1 + \tau_3 + \dots + \tau_{2k+1})}dt 
  \right\rangle,
\end{align}
where $\tau_k$ is the $k$th sojourn time, and thus $t_k = \sum_{i=1}^{k} \tau_i$.
Integrating the above equations, and using interindependence of $\tau_k$ and
$\tau_l$ ($k\neq l$), we have 
\begin{align}
  \hat{g}^{+}_{2k+1}(u;s)
  &=
  \frac{1- \hat{\rho}_-(s)}{s} \hat{\rho}_-^k(s) \hat{\rho}_+^{k+1}(s+u),
  \\[0.1cm]
  \hat{g}^{+}_{2k}(u;s) 
  &= \frac{1- \hat{\rho}_+(s+u)}{s+u} \hat{\rho}_-^k(s) \hat{\rho}_+^{k}(s+u).
\end{align}
The cases in which the system starts from $-$ state can be calculated in 
the similar way. 
Then, the the PDF of $T_+(t)$ is obtained by summing up $g^{\pm}_n(y;t)$ in
terms of $n$: ${g}^\pm (y;t) = \sum_{n=0}^\infty g^\pm_n (y;t)$, and thus we
have
\begin{align}
  \hat{g}^+ (u;s)
  &=
  \frac{1-\hat{\rho}_-(s)}{s \hat{\rho}(s,u)} \hat{\rho}_+(s+u) +
  \frac{1-\hat{\rho}_+(s+u)}{(s+u)\hat{\rho}(s,u)},
  \\[0.1cm] 
  \hat{g}^-(u;s) 
  &=
  \frac{1-\hat{\rho}_+(s+u)}{(s+u)\hat{\rho}(s,u)} \hat{\rho}_-(s) +
  \frac{1-\hat{\rho}_-(s)}{s\hat{\rho}(s,u)},  
\end{align}
where $\rho(s,u)\equiv 1- \hat{\rho}_+(s+u)\hat{\rho}_-(s)$.
In the small $s$ and $u$ limit, 
\begin{align}
  \hat{g}^{\pm}(u;s)  
  &\sim
  \frac{1-\hat{\rho}_-(s)}{s \hat{\rho}(s,u)}  +
  \frac{1-\hat{\rho}_+(s+u)}{(s+u)\hat{\rho}(s,u)}.
  \label{Laplace_pdf_occupation}
\end{align}

\section {Distributional limit theorems}


%
%

\subsection{ Case (1)} 
From Eq.~(\ref{Laplace_pdf_occupation}), the Laplace transform
of the PDF of $T_+(t)$ for the case (1) is given by
\begin{equation}
  \hat{g}^\pm(u;s) 
  \sim
  \frac{a_-s^{\alpha-1} + \mu_+}{a_-s^{\alpha} + \mu_+(s+u)}.
\end{equation}
Using the relation between the moments of $T_+(t)$ and $\hat{g}^\pm(u;s)$, we
have the asymptotic behavior of the $n$th moment of $T_+(t)$
\begin{align}
  \label{moments_case1}
  \left\langle T_+^n(t) \right\rangle_{\pm}
  &\sim
  \left(\frac{\mu }{a_-}\right)^{\!\!n}\!\!
  \frac{n! \,t^{n\alpha}}{\Gamma(1 + n\alpha)},
\end{align}
where $\mu=m_1$. 
It follows that the ETMSD shows normal diffusion:
\begin{equation}
\langle \overline{\delta^{2}(\Delta;t)}  \rangle 
\sim 2n \left[D_- + \frac{\mu (D_+ - D_-)}{a_-\Gamma(1+\alpha)} \frac{1}{t^{1-\alpha}} \right] \Delta,
\end{equation}
where we used $\langle \overline{D_\pm(t)}\rangle \sim D_\pm$ and Eq.~(\ref{moments_case1}). 
Because TMSD converges to $2nD_- \Delta$ as $t\to \infty$, this process seems to be normal diffusion.

In Brownian motion, $\overline{D(t)}$ converges to a constant and the distribution follows Gaussian. Thus, deviation from Gaussian detects anomaly 
of the process. Since $\overline{D(t)}$ is given by Eq.~(\ref{D_t_TAMSD}) and $\overline{D(t)} \to D_-$,
 we consider the deviation, i.e., $\delta D_t \equiv \overline{D(t)} - D_-$.
 By  Eq.~(\ref{D_t_TAMSD}), we have 
 \begin{equation}
\frac{\delta D_t}{\langle \delta D_t\rangle} \cong \frac{(\overline{D_-(t)} - D_-)t}{(D_+ - D_-)\langle  T_+(t)\rangle} +  \frac{T_+(t)}{\langle  T_+(t)\rangle}.
\label{relation_delta_Dt_and_Tt}
 \end{equation} 
 Here, the first term in the right-hand side can be neglected if $\langle (\overline{D_-(t)} - D_-)^2t^2 \rangle 
 =o(\langle  T_+(t)^2\rangle - \langle T_+(t)\rangle^2$). Note that this condition is satisfied when $\alpha >0.5$ [see Eq.~(\ref{moments_case1})]. 
By Eq.~(\ref{moments_case1}), moments of the normalized occupation time defined by $T_\alpha(t) \equiv T_+(t) /\langle T_+(t) \rangle$ 
becomes
\begin{equation}
\langle T_\alpha(t) ^n \rangle \sim \frac{n! \Gamma(1+\alpha)^n}{\Gamma(1 + n\alpha)}\quad (t\to\infty). 
\end{equation}
When the PDF of a random variable $M_\alpha$ follows the Mittag-Leffler distribution of order $\alpha$, 
the Laplace transform is given by
$
\langle e^{-z M_\alpha} \rangle = \sum_{k=0}^\infty \frac{\Gamma(1+\alpha)^k z^k}{\Gamma(1+k\alpha)}.
$
 Therefore, 
 the distribution of  $\delta D_t/\langle \delta D_t\rangle$ is not Gaussian but 
 converges  to the Mittag-Leffler distribution when $\alpha>0.5$ (see Fig.~\ref{limit_dist_case2}a). For $\alpha<0.5$, the first term in 
 Eq.~(\ref{relation_delta_Dt_and_Tt}) becomes the leading term and the distribution of $\delta D_t/\langle \delta D_t\rangle$ becomes Gaussian 
 with the mean 0 and the variance $2 \Delta D_-^2 a_-^2\Gamma(1+\alpha)^2 t^{1-2\alpha}/\{n(D_+ - D_-)^2\mu^2\}$. 
 For $\alpha > 0.5$, using Eq.~(\ref{rsd_occupation_time}) yields
 \begin{equation}
 \Sigma^2 (t;\Delta) \sim \frac{\mu^2 (D_+ - D_-)^2 A(\alpha)}{a_-^2 D_-^2 \Gamma(1+\alpha)^2}
  t^{-(1-\alpha)},
 \end{equation}
 where $A(\alpha)=\frac{2\Gamma(\alpha+1)^2}{\Gamma(2\alpha+1)} -1$.
 For $D_-=0$, the result is exactly the 
same as that in CTRW \cite{He2008}:  
$
\Sigma(t;\Delta) \sim \sqrt{A(\alpha)}.
$

%
%

\subsection{ Case (2)} 
In this case, Eq.~(\ref{Laplace_pdf_occupation}) yields the Laplace transform of the PDF of $T_+(t)$: 
\begin{equation}
\hat{g}^\pm(u;s) 
\sim \frac{a_+(s+u)^{\alpha_+-1}+a_-s^{\alpha-1}}{a_+(s+u)^{\alpha_+} +a_-s^{\alpha}}.
\end{equation}
The Laplace transform of the first moment $\langle T_+(t)\rangle$ is scaled as 
\begin{equation}
\langle \hat{T}_+(s)\rangle = \left. \frac{\partial \hat{g}^\pm(u;s)}{\partial u}\right|_{u=0} \sim 
- \frac{a_+ }{a_-} \frac{1}{s^{2 - \delta-\alpha}},  
\end{equation}
where $\delta \alpha = \alpha_+ - \alpha$. Thus, 
The asymptotic behavior of $\langle T_+(t)\rangle$ becomes 
\begin{equation}
\langle T_+(t)\rangle  \sim  \frac{a_+}{a_-\Gamma(2 -\delta \alpha)} t^{1-\delta \alpha}, 
\label{mean_occupation_2b}
\end{equation}
Moreover, the second moment of $T_+(t)$ is scaled as 
\begin{equation}
\langle T_+(t)^2 \rangle \sim \frac{2a_+ (1-\alpha_+)}{a_- \Gamma(3- \delta \alpha)} t^{2- \delta \alpha}.
\end{equation}
It follows that the second moment of $T_+(t)/\langle T_+(t) \rangle$
diverges for $t\to \infty$. 
Using Eqs.~(\ref{D_t_TAMSD}) and (\ref{mean_occupation_2b}) yields the ETMSD:
\begin{equation}
\langle \overline{\delta^{2}(\Delta;t)}  \rangle 
\sim 2n \left[D_- + \frac{a_+ (D_+ - D_-)}{a_-\Gamma(2-\alpha_+ + \alpha)} \frac{1}{t^{\alpha_+ -\alpha}} \right] \Delta.
\end{equation}
As in the previous case, TMSD converges to $2nD_- \Delta$ as $t\to \infty$. By  Eq.~(\ref{rsd_occupation_time}), 
the RSD decays as 
\begin{equation}
\Sigma^2(t;\Delta) \sim \frac{2a_+(D_+ - D_-)^2 (1-\alpha_+)}{a_- D_-^2 \Gamma(3-\delta \alpha)} t^{- \delta \alpha}. 
\end{equation}

\begin{figure}
\includegraphics[width=1.\linewidth, angle=0]{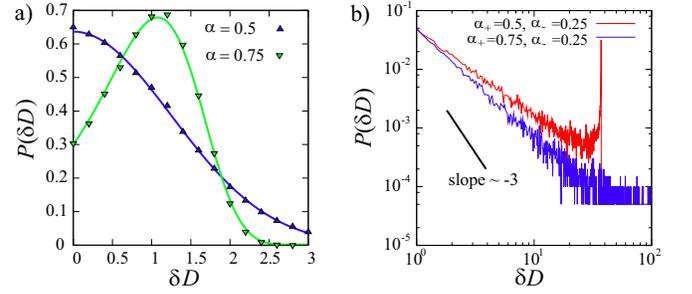}
\caption{Distribution of deviations of the time-averaged diffusivity, $\delta D\equiv \delta D_t / \langle D_t \rangle$, 
in the cases (1) and (2), corresponding to (a) and (b), respectively ($D_-=1$, $D_+=10$, and $t=10^4$).
In Fig.~(a), the Mittag-Leffler distributions are drawn by solid lines. 
In Fig.~(b), power-law distribution with exponent $-3$ is drawn for reference.  
 Squares with colors are the results of numerical 
simulations. 
In numerical simulations, we used the following power-law distribution 
for sojourn time distribution: $\rho_\pm (\tau) = c_\pm \tau^{-1-\alpha_\pm}$ for $\tau\geq \tau_0^\pm$ 
and $0$ for $\tau<\tau_0^\pm$, where $c_\pm$ is the normalization constant. In case (1), we used the 
exponential distribution for $\rho_+(\tau)$.
 }
\label{limit_dist_case2}
\end{figure}

Although we do not have the limit distribution of $T_+(t)/ \langle T_+(t) \rangle$, the tail should be a heavy tail (power-law 
distribution) because the second moment of  $T_+(t)/ \langle T_+(t) \rangle$ diverges. By the relation between 
$\delta D_t$ and $T_+(t)$, i.e., Eq.~(\ref{relation_delta_Dt_and_Tt}), we find that  
the deviations of time-averaged diffusion coefficient, $\delta D_t/\langle \delta D_t\rangle$, are random and the 
distribution is a non-trivial distribution characterized by a power law (see Fig.~\ref{limit_dist_case2}b).
This is a similar situation for the PDF of time-averaged diffusion coefficients in some parameter region of SEDLF \cite{Akimoto2014}.

%
%

\subsection{ Case (3)}
\begin{figure}
\includegraphics[width=1.\linewidth, angle=0]{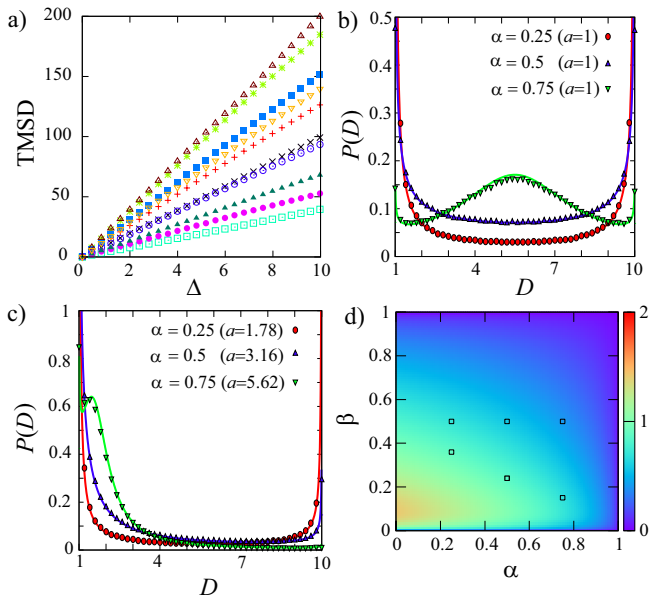}
\caption{Anomalous fluctuations of time-averaged diffusivity in the case 3 ($D_-=1$ and $D_+=10$).
(a) TMSDs for 10 different realizations ($\alpha=0.5$ and $t=10^4$).
(b) and (c) Distribution of time-averaged diffusion coefficients ($t=10^3$).
Symbols are the results of numerical simulations and solid curves are the theoretical ones. 
(d) RSD as a function of $\alpha$ and $\beta$ ($t=10^5$). Squares with colors are the results of numerical 
simulations. In numerical simulations, we used the same power-law distribution as in Fig.~\ref{limit_dist_case2}.
 }
\label{limit_dist_case3}
\end{figure}

Contrary to the previous two cases, TMSDs do not converge to a constant in the case (3), whereas 
TMSD shows normal diffusion [see Eq.~(\ref{e.tmsd.approx})].
Eq.~(\ref{Laplace_pdf_occupation}) yields the Laplace transform of the PDF of $T_+(t)$: 
\begin{equation}
\hat{g}^\pm(u;s)  
\sim \frac{a_+(s+u)^{\alpha-1}+a_-s^{\alpha-1}}{a_+(s+u)^{\alpha} +a_-s^{\alpha}}.
\label{Laplace_case3}
\end{equation}
By Appendix~B in \cite{God2001}, Eq.~(\ref{Laplace_case3}) implies that the limit distribution of $T^+(t)/t$ 
exists:  
\begin{equation}
\lim_{t\to\infty}  g_{T^+/t}(x) = g_{\alpha,\beta}(x), 
\end{equation}
and the distribution is given by 
\begin{equation}
g_{\alpha,\beta}(x) =
 \frac{(a \sin \pi \alpha/\pi) x^{\alpha-1} (1-x)^{\alpha-1}}{a^2 x^{2\alpha} +2a \cos \pi \alpha (1-x)^\alpha x^\alpha + (1-x)^{2\alpha}},
\label{T+_limit_dist}
\end{equation}
where $g_{T^+/t}(x)$ is the PDF of $T_+(t)/t$, $a=a_-/a_+$ and $\beta\equiv 1/(1+a)$. 
This is the Lamperti's generalized arcsine law \cite{Lamperti1958}, which is observed for time-averaged drift in superdiffusion \cite{Akimoto2012}. 
By Eq.~(\ref{D_t_TAMSD}),  the distribution of the time-averaged diffusion coefficient is given by that of $T_+(t)/t$:
\begin{eqnarray}
\Pr \left(\overline{D(t)}\leq x\right) 
&=& \Pr \left(  \frac{T_+(t)}{t} \leq  \frac{x-D_-}{D_+-D_-}\right). 
\end{eqnarray}
Because the PDF of $T_+(t)/t$ follows the Lamperti's generalized arcsine law, Eq.~(\ref{T+_limit_dist}), the PDF of $\overline{D(t)}$ is given by
$
P_D(x) = g_{\alpha,\beta} \left(\frac{x-D_-}{D_d}\right)/D_d,
$ 
 where $D_d=D_+ - D_-$.
Because the mean and second moment of $T_+(t)/t$ are given by $\langle T_+(t)/t \rangle = \beta$ and 
$\langle (T_+/t)^2 \rangle = m(\alpha,\beta)\equiv \beta (\alpha \beta +1 - \alpha)$, respectively \cite{God2001}, we have the RSD
\begin{equation}
\Sigma(t;\Delta) \sim \sqrt{\frac{D_-^2 + 2D_- D_d\beta+D_d^2 m(\alpha,\beta)}{(D_- + D_d\beta)^2} -1}.
\end{equation}
As shown in Fig.~\ref{limit_dist_case3}, theory is in good agreement with numerical results.

\section{ Conclusion}
We have shown three distributional limit theorems for time-averaged observables related to diffusivity in 
the Langevin equation with dichotomously fluctuating diffusivity. When one of the states is zero ($D_-=0$)
 in the case (1), statistical properties of TMSD are exactly the same as those in CTRW. 
Therefore, this model is a generalization of CTRW. When both diffusion coefficients are not zero, 
the TMSD asymptotically show normal diffusion in all cases, whereas fluctuations of  TMSD
(deviations of time-averaged diffusion coefficients) are intrinsically random. 
Especially in case (3), time-averaged diffusion coefficients are intrinsically random and the distribution 
follows the generalized arcsine law. 
As a result, we have found anomalous fluctuations in apparently normal diffusion processes.

\section*{ Acknowledgments}
We thank T. Miyaguchi and T. Uneyama for discussions on these issues. 
T. A. was partially supported by a Grant-in-Aid for Young Scientists (B) (26800204). 
E. Y. was funded by MEXT Grant-in-Aid for the ``Program for Leading Graduate Schools."



\appendix
\section{Derivation of Eq.~(\ref{rsd_occupation_time2})}
Here, we derive the RSD in Brownian motion with the diffusion coefficient $D_-$. Since this process is described by Brownian motion, displacement 
  $\delta \bm{r}(\Delta;t) \equiv \bm{r}(\Delta + t) - \bm{r}(t)  $ follows a Gaussian distribution with the mean 0 and the variance $2nD_- \Delta$. 
  The mean TMSD is straightforwardly calculated as $\langle \{\overline{\delta^{2}(\Delta;t)}\} \rangle= 2nD_- \Delta$.
The second moment of TMSD can be calculated as follows:
\begin{widetext}
\begin{align}
\langle \{\overline{\delta^{2}(\Delta;t)}\}^2 \rangle
&\sim \frac{2}{t^2} \int_0^t dt' \int_{t'}^t dt'' \langle \delta \bm{r}^2(\Delta;t') \delta \bm{r}^2(\Delta;t'') \rangle\\
&= \frac{2}{t^2} \int_0^t dt' \int_{t'}^{t'+\Delta} dt'' \langle \delta \bm{r}^2(\Delta;t') \delta \bm{r}^2(\Delta;t'')  \rangle
+ \frac{2}{t^2} \int_0^t dt' \int_{t'+\Delta}^t dt'' \langle \delta \bm{r}^2(\Delta;t')  \rangle \langle  \delta \bm{r}^2(\Delta;t'')  \rangle\\
&= \frac{2}{t^2} \int_0^t dt' \int_{t'}^{t'+\Delta} dt''\left\{ \langle \delta \bm{r}^2(t'' -t';t') \rangle \langle \delta \bm{r}^2(\Delta;t'') \rangle 
+ \langle \delta \bm{r}^4(t'+\Delta - t'';t'') \rangle \right.\\
&+ \langle  \delta \bm{r}^2(t'+\Delta-t'';t'') \rangle \langle \delta \bm{r}^2(t''-t';t'+\Delta) \rangle \}
+\frac{2}{t^2} \int_0^t dt' (t-t'-\Delta) (2nD\Delta)^2\\
&=(2nD \Delta)^2 \left( 1 + \frac{4\Delta}{3nt}  \right).
\end{align}
\end{widetext}
It follows that the RSD decays as
\begin{equation}
\Sigma^2(t;\Delta) 
\sim \frac{4\Delta}{3nt} \quad (t \to \infty ).
\label{EB}
\end{equation}
 
%


\end{document}